\documentclass[fleqn,usenatbib]{mnras}
\usepackage{newtxtext,newtxmath}
\usepackage[T1]{fontenc}
\usepackage{ae,aecompl}
\usepackage{graphicx}	
\usepackage{amsmath}	
\usepackage{amssymb}	
\usepackage{epsf}
\usepackage{epstopdf}
\usepackage {amssymb}
\usepackage{multirow}
\newcommand{\nc}{\newcommand}
\nc{\ba}{\begin{eqnarray}}
\nc{\ea}{\end{eqnarray}}
\newcommand\be{\begin{equation}}
\usepackage{float}
\newcommand\ee{\end{equation}}

\usepackage{placeins}

\nc{\e}{{\bf{e}}}
\usepackage{hyperref}
\usepackage{cleveref}
\usepackage{pdflscape}

\title{Probing Intermediate Mass Black Holes in M87 through Multi-Wavelength Gravitational Wave Observations}

\author[Razieh Emami and Abraham Loeb]{
Razieh Emami
and Abraham Loeb
\\
Center for Astrophysics, Harvard-Smithsonian, 60 Garden Street,  Cambridge, MA 02138, USA
}
\date{Accepted XXX. Received YYY; in original form ZZZ}

\pubyear{2019}

\begin{document}
\label{firstpage}
\pagerange{\pageref{firstpage}--\pageref{lastpage}}
\maketitle

\begin{abstract}
We analyze triple systems composed of the super massive black hole (SMBH) near the center of M87 and a pair of black holes (BHs) with masses in the range $10-10^3$ $M_{\odot}$. We consider the post Newtonian precession as well as the Kozai-Lidov interactions at the quadruple and octupole levels in modeling the evolution of binary black hole (BBH) under the influence of the SMBH. Kozai-Lidov oscillations enhance the gravitational wave (GW) signal in some portions of the parameter space. We identify frequency peaks and examine the detectability of GWs with LISA as well as future observatories such as $\mu$Ares and DECIGO. We show examples in which GW signal can be observed with a few or all of these detectors. Multi-Wavelength GW spectroscopy holds the potential to discover stellar to intermediate mass BHs near the center of M87. We estimate the rate, $\Gamma$,  of collisions between the BBHs and flyby stars at the center of M87. Our calculation suggest $\Gamma < 10$ $\rm{Gyr}^{-1}$ for a wide range of the mass and semi-major axes of the inner binary. \\ 
\end{abstract}


\section{Introduction}

Black holes (BHs) are ubiquitous in galaxies and form across a wide range of masses and environments. Stellar-mass BHs are most abundant, with observational evidence first detected in X-rays \citep{1972Natur.235...37W, 2006ARA&A..44...49R}, and more recently by gravitational waves (GWs) with LIGO/Virgo \citep{2016PhRvL.116f1102A, 2016PhRvL.116x1103A}. Intermediate mass black holes (IMBHs) are expected to form from the accretion of gas in dwarf galaxies (see \citep{2019arXiv190904670R} and references therein), mergers of stars in dense stellar clusters \citep{1999A&A...348..117P, 2009ApJ...694..302D, 2012MNRAS.425L..91P, 2016MNRAS.459.3432M}, from direct collapse of inflowing dense gas in protogalaxies \citep{1994ApJ...432...52L}, collapse of Pop III stars from early universe \citep{2001ApJ...551L..27M, 2004ARA&A..42...79B}, from supermassive stars in AGN accretion disk instabilities \citep{2012MNRAS.425..460M, 2014MNRAS.441..900M} or as recently proposed by \cite{2015MNRAS.454.3150G} a consequence of dynamical interaction between hard binaries, containing stellar mass BH, and other stars and binaries.

Finding unambiguous observational evidence for IMBH candidates in galactic nuclei is challenging due to the short lifetime associated with mergers and accretion by the supermassive black holes (SMBHs) there \citep{2014GReGr..46.1702N,2016PASA...33....7J}. 
There is however some evidence for their existence in the local Universe \citep{2017IJMPD..2630021M}. In particular, Ultraluminous 
X-ray sources (ULXs) imply BHs with masses above 20 $M_{\odot}$ in some cases \citep{2004IJMPD..13....1M}. A recent discovery of IMBH candidates in 
dwarf galaxies at ($z \lesssim 2.4 $) was reported in the Chandra COSMOS Legacy Survey \citep{2018MNRAS.478.2576M}, using a sample of 40 active galactic nuclei (AGN) in dwarf galaxies with redshift below 2.4. More recently \citep{2018ApJ...863....1C} used data mining in wide-field sky surveys and identified a sample of 305 IMBH candidates in galaxy centers, accreting gas which creates characteristic signatures of a type I AGN. Most recently, a new sample of Wandering IMBH was discovered using the large-scale and high resolution radio telescopes 
\citep{2019arXiv190904670R} in nearby dwarf galaxies. 

New probe of IMBH candidates are awaited using 30-m class telescopes \citep{2019arXiv190308670G}.
IMBH candidates may also be able to detect using the GW signals in different frequencies \citep{2012PhRvD..85l3005K}.

SMBHs, with masses in the range $10^6-10^{10} M_{\odot}$ are believed to exist in the core of almost all of massive galaxies \citep{2015ApJ...807L...9W, 2015ApJ...805..179B}. This includes SgrA* at the Galactic center \citep{1998ApJ...509..678G,2018A&A...618L..10G}, as well as the nearby 
elliptical galaxy M87 \citep{2009ApJ...700.1690G, 2011ApJ...729..119G,2013ApJ...770...86W}. Most recently 
the mass of the M87 SMBH was precisely measured by Event Horizon Telescope (EHT) collaboration \citep{2019ApJ...875L...1E,2019ApJ...875L...2E, 2019ApJ...875L...3E, 2019ApJ...875L...4E,2019ApJ...875L...5E,2019ApJ...875L...6E} to be $ 6.5 \times 10^9 M_{\odot}$. 
In addition to the above constraints on the mass of M87, there are some studies on the proper motion and core stability in Virgo cluster. \cite{2009nsem.confP...3W} monitored the relative position of M87 and M84 at 43 \rm{GHz} with the Very Long Baseline Array Astrometry(VLBA). This study demonstrated a 5 sigma detection of a relative motion $~ 800$ km/s.

Being at the center of the virgo cluster, M87 is the nearest giant elliptical galaxy which is the
product of many galaxy mergers. As a result, M87 should contain a large population of stellar mass BHs and possibly a handful of IMBH candidates which used to be at the center of dwarf galaxies that merged in. Here we propose to search for this IMBH populations in M87 through their GW signals. 

A large fraction of the IMBH candidates in M87 might result in binaries in eccentric orbits around the SMBH. 
This system is well described by the hierarchical triple approach \citep{2013ApJ...773..187N,2017MNRAS.466..276G, 2018MNRAS.481.4907G,2019arXiv190208604R,2019ApJ...875L..31H,2019MNRAS.486.5008A} in which the inner binary contains of IMBH candidates while the outer binary includes the SMBH. The IMBH candidates in the inner binary are expected to emit bursts of GWs at any pericenter passage. The frequency of these GW bursts depend on the orbital parameters.

In this paper, we simulate triple systems composed of M87 and a pair of the BHs and consider the dynamical evolution of the binary BHs. The induced Kozai-Lidov oscillations \citep{2013ApJ...773..187N,2018MNRAS.480L..58A, 2018ApJ...863....7R,2019arXiv190208604R,2019ApJ...875L..31H,2019MNRAS.486.5008A} are ubiquitous for low mass BHs. Increasing the mass of BHs in the binary suppresses the strength of the Kozai-Lidov oscillations. We present examples in which the GW signal is above the noise of future observatories, including LISA \citep{2019CQGra..36j5011R, 2019arXiv190302579E}, $\mu$-Ares \citep{2019arXiv190811391S}, a newly proposed space based GW mission with the ability of filling the gap between the milli-Hz and nano-Hz frequency windows surveyed by LISA , and also Decihertz Observatories (DOs) such as the Decihertz Interferometer GW observatory (DECIGO) (e.g. \cite{2019arXiv190811375A} and Refs. in this). Adapting a maximum observational time of up to 10 \rm{yrs}, we present examples with detectable GW signals
by these observatories. 

The paper is organized as follows. In Sec. \ref{Triple-Detail} we briefly review the details of the related hierarchical triple system. In Sec. \ref{stability} we introduce the stability conditions that must be taken into account.
In \ref{GW-Triple} we review the finite time Fourier transformation for computing the GW signal. In Sec. \ref{Examples} we present a variety of different examples with a potentially detectable GW signal as demonstrated in Sec. \ref{Detection}.
In Sec. \ref{non-equal} we briefly consider binaries with non-equal BH masses. In Sec. \ref{Collision-Rate} we estimate the rate of collision between the inner BBH and the flyby stars. 
In Sec. \ref{Lifetime} we compute the lifetime of the related systems.
we discuss about the detectability of the GW signal. Finally, we summarize our conclusions in Sec. \ref{Conc}. 

\section{Impact of SMBH on the evolution of BBHs}
\label{Triple-Detail}
We model the dynamical influence of a SMBH on the evolution of BBHs, accounting for inner pericenter precession at first order, quadruple and octupole terms in the Lidov-Kozai interaction. 

The full Hamiltonian of the triple system is given by,

\ba
\label{total-H}
H_{tot} = H_{LK} + H_{1PN},
\ea
where $H_{LK}$ and $H_{1PN}$ present the Lidov-Kozai Hamiltonian and first order post Newtonian precession, respectively. $H_{LK} = H_1 + H_2 + H_{12}$ with $H_{i}, i = 1,2$ referring to the Keplerian Hamiltonian for the inner and outer binaries in the system and $H_{1,2}$ denoting the interaction term between them. The interaction term is expressed as a series expansion in the separation of two binaries. We make use of Eqs. (5-8) of \cite{2018MNRAS.480L..58A,2018ApJ...863....7R} for modeling the above components of full Hamiltonian. 

Using Eq. (\ref{total-H}) we derive the equations of motion for the inner and outer semi-major axes. We also take into account GW emission in the inner orbit as an extra term that shrinks the orbit of the BHs. 

\section{Stability Conditions}
\label{stability}
Here we present some stability conditions that must be taken into account in our analysis:

\begin{itemize}
    \item  The orbital parameters must be selected such that prevent the inner binary from reaching the Roche limit of the outer-binary \citep{2018ApJ...856..140H, 2019MNRAS.488...47F}. This implies, 
    \ba 
    \label{Roche-Lobe}
     \frac{a_{out}}{a_{in}} > \left( \frac{1 + e_{in}}{1 - e_{out}}\right) \bigg{(} \frac{3 M_{SMBH}}{m_1+m_2}\bigg{)}^{1/3}.
    \ea
 The above widely used analytical stability criteria does not account for possible changes in the Hill radius due to the inclination \citep{2017MNRAS.466..276G}. However, as it was demonstrated  in \citep{2017MNRAS.466..276G}, the numerical fit gives us an excellent agreement to the analytical expression up to 120 degree and a marginal agreement afterward.
    \item The system must possess dynamical stability, implying that the hierarchical secular treatment is satisfied \citep{2014ApJ...795..102N,2018ApJ...856..140H},
    \ba
    \label{dynamical}
    \left(\frac{a_{in}}{a_{out}}\right) \left(\frac{e_{out}}{1 - e^2_{out}}\right) < 0.1.
    \ea
    \item For eccentric outer orbits, we add one extra criteria for the pericenter of the outer orbit. We urge the pericenter distance to be relatively larger than the event horizon of SMBH,
\ba 
\label{Event-BH}
   a_{out} \left( 1 - e_{out} \right) > 2 \left( \frac{G M_{SMBH}}{c^2}\right).
\ea
This criteria makes us ensure that the outer orbit does not hit the event horizon of the SMBH. This is expected as otherwise we lose the outer binary at the first pericenter crossing. While this criteria is easier to achieve for moderately massive SMBH, such as SgrA*, it needs to be checked for a massive SMBH like M87.

\end{itemize}

\section{Gravitational Waves Estimation}
\label{GW}
Given the dynamical evolution of the orbital parameters, we calculate the GW amplitude from eccentric binary black holes (hereafter EBBHs) and study the detectability of their GW signal. Using this formalism, we present some examples with potentially detectable GW signals in M87. 

\begin{figure*}
\includegraphics[height=160pt,width=500pt,trim = 6mm 1mm 0mm 1mm]{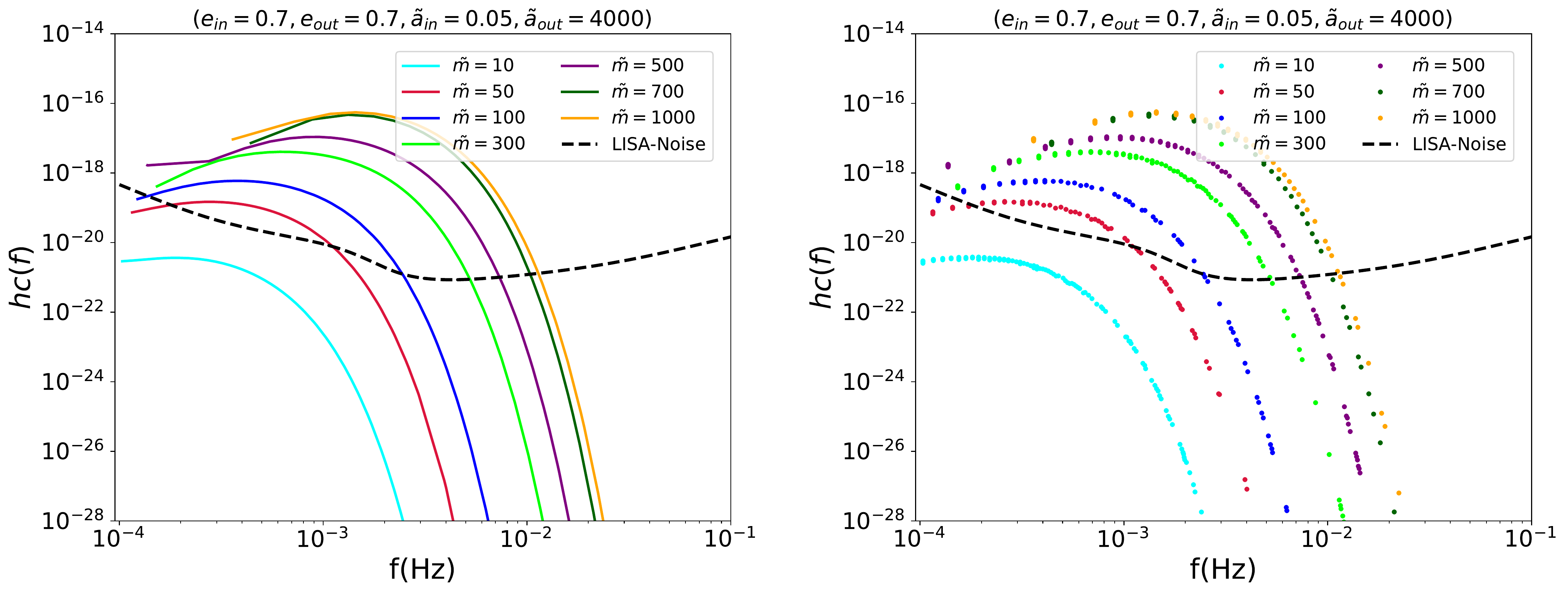}
\caption{GW signal (in color) relative to the LISA noise (dashed black line) for equal mass binaries with individual masses $ \tilde{m} \equiv m/ M_{\odot} = (10, 50, 100, 300, 500, 700, 1000) $. Left: the interpolated GW signal with $\pi (f - f_n) \Delta T <1$. Right: discrete frequencies with the above condition. The Signal to Noise ratio for LISA in the above examples is \rm{S/N} = (0.1, 0.8, 5.7, 82, 224), respectively.}
\label{mass-points}
\end{figure*}

\subsection{GW-Computation}
\label{GW-Triple}
Unlike binaries on circular orbits, EBBHs emit GW in a discrete spectrum \citep{1963PhRv..131..435P} with characteristic frequencies $f_{n} = n f_{\rm{orb}}$, where $n$ refers to the harmonic index while $ f_{\rm{orb}} =  (2\pi)^{-1} \sqrt{G(m_1 + m_2)}  a^{-3/2}$ is the orbital frequency in a circular orbit with semi-major $a$ and $m_1$ and $m_2$ refer to the mass of BHs in the inner binary. 
The amplitude of the GW signal is given by, 
\ba 
\label{haet}
h(a,e,t) = \sum_{n = 1}^{\infty} h_n(a,e, f_n) \exp{\left(2 \pi i f_n t\right)},
\ea
where $h_n(a,e, f_n)$ is defined as,

\ba 
\label{hane}
h_n(a,e, f_n) = \frac{2}{n} \sqrt{g(n,e)} h_0(a),
\ea
with $h_0(a)$ being the dimensionless strain for a circular orbit, given by  \citep{1963PhRv..131..435P,2019arXiv190208604R, 2019ApJ...875L..31H},
 
\ba 
\label{h-cir}
h_0(a) = \sqrt{\frac{32}{5} } \frac{G^2}{c^4} \frac{m_1 m_2}{D a},
\ea
and where $D$ denotes the angular diameter distance to the source. 

To gauge the detectability of the GW signal, we divide the stream of GW into time intervals with a duration $\Delta T$ and 
perform a Finite Fourier Transformation (FFT) of the GW strength for the duration $\Delta T $. The Fourier component of the GW signal can be computed by taking a time integral of Eq. (\ref{haet}) from $-(\Delta T/2)$ to $(\Delta T/2)$ yielding, 

\begin{figure*}
\includegraphics[height=160pt,width=500pt,trim = 6mm 1mm 0mm 1mm]{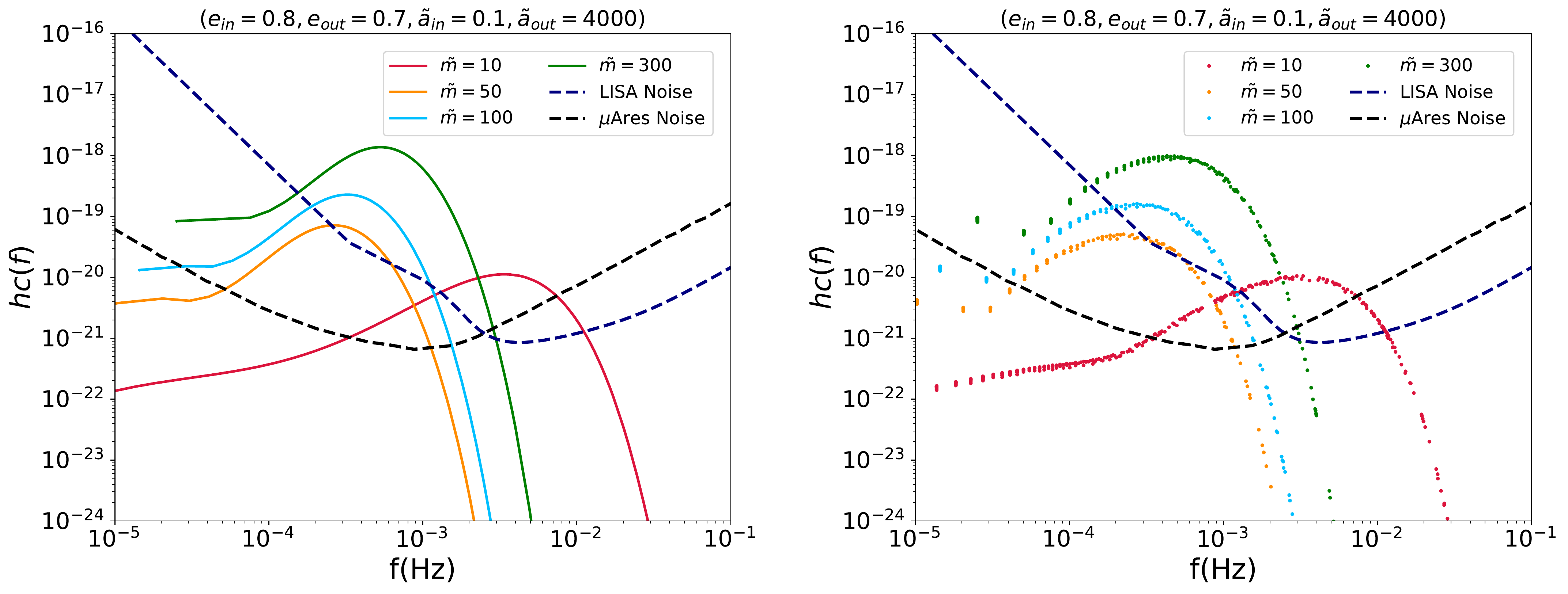}
\caption{GW signal (in color) compared to $\mu$-Ares (dashed blue) and LISA noise (dashed black) for BBHs with individual mass $ \tilde{m} = (10, 50, 100, 300) $. Left: the interpolated GW signal with $\pi (f - f_n) \Delta T <1$. Right: discrete frequencies with the above condition. Signal to Noise ratios for the $\mu$-Ares examples are \rm{S/N} = (91, 266, 947, 7.5 $\times 10^3$).}
\label{mass-points-Mu-Ares}
\end{figure*}
\ba 
\label{Fourier-Transformation}
\tilde{h}(a,e,f) = \sum_{n =1}^{\infty} h_n(a,e; f_n) \Delta T w(f, f_n, \Delta T),
\ea
where,
\ba 
\label{waef}
w(f, f_n, \Delta T) = \frac{\sin{\left[ \pi \left(f - f_n \right) \Delta T\right]}}{\pi \left(f - f_n \right) \Delta T}.
\ea
The observational interval $\Delta T$ plays an important role. For close-in sources, the strength of GW is large enough to allow smaller values of $\Delta T$. This leads to wider frequency bins, as $f -f_n \simeq 1/\Delta T$. The situation is different for wide-separation
binary systems, where the observational time $\Delta T$ must be larger. 
Therefore the GW signal is localized in the frequency domain around specific harmonics. We define \textit{characteristic} detectable frequencies which are aimed to be within the frequency range of LISA or ground based GW detectors,
\ba 
\label{characteristic-nu}
f_{min} \leq f \equiv f_{n} \pm \alpha/\Delta T \leq  f_{max},
\ea
where $\alpha \lesssim 1$ and with $f_{n} = n f_{\rm{orb}}$. \\
Next, we define the characteristic strain of the GW signal as, 
\ba 
\label{hc}
h_c^2(a, e, f) \equiv 4 f^2 |\tilde{h}(a,e,f)|^2.
\ea
To check the detectability of the GW signal, Eq. (\ref{hc}) must be compared with the 
characteristic noise of GW observatories. As discussed in the introduction, we will consider several different GW detectors, including LISA, $\mu$-Ares and DECIGO. 

Defining the noise generically as $S_{c}(f)$, the signal to noise ratio (S/N) is given by, 

\ba 
\label{signal-noise}
(\rm{S/N})^2(a, e) = \int_{f_{min}}^{f_{max}} \frac{h_c^2(a, e, f)}{S_c(f)} d(\ln{f}).
\ea
\subsection{GW signal in M87}
\label{Examples}
Having presented a formalism to compute the GW signal for a generic system, we now apply it to the case of M87. Since $D = 16 \rm{Mpc}$, we choose a relatively large value for the observational time, $\Delta T$. From Eq. (\ref{Fourier-Transformation}), a large $\Delta T$ has two different effects, though. On the one hand, it enhances the GW amplitude. On the other hand, it diminishes the frequency width of the GW signal as $(f - f_n) \simeq 1/\Delta T$. This leads to a discrete spectrum of observable frequencies. The combination of these two effects lead to a range of $\Delta T$ that can lead to an observable GW signal at a range of discrete frequencies.

To clarify the above points, we present some examples with potentially detectable GW signals at different frequencies. We choose $\Delta T $ differently in each of these examples to both help pushing the GW signals above the noise as well as increasing the amount of observable frequencies. 

Throughout our analysis, we neglect the GW bursts from the inspiral phase of IMBH around M87. In another paper \citep{2019arXiv190302579E}, we estimated the timescale associated with these bursts to be, 

\ba 
\label{GW-Inspiral}
&\tau_{GW} \simeq \left(\frac{5}{64}\right) \frac{c^5 a^4_{out} \left( 1 - e^2_{out} \right)^{7/2}}{G^3 (m_1 + m_2) M^2_{87}} \left(1 + \frac{73}{24} e^2_{out} + \frac{37}{96}  e^4_{out} \right)^{-1}. \nonumber\\
&&
\ea
It is easy to see that $\tau_{GW} \simeq \mathcal{O}(\rm{Gyr})$ is much longer than both of
the observational time and the lifetime of the orbit of IMBH. Therefore, we can safely ignore the GW decay time in our analysis. 


\subsubsection{Probing GW in LISA}
First, we consider GWs in the LISA band. We adopt the LISA noise curve \citep{2019CQGra..36j5011R, 2019arXiv190302579E} and present the characteristic GW signal on the top of that. 
Figure \ref{mass-points} shows the GW signal in the LISA band for few different equal mass binaries with each component having, $\tilde{m} (\equiv m/M_{\odot}) = (10, 50, 100, 300, 500, 700, 1000)$. The rest of the parameters are chosen to be the same: $\tilde{a}_{in} = 0.05, \tilde{a}_{out} = 4000, e_{in} = 0.7, e_{out} = 0.7$, where hereafter $\tilde{a} \equiv (a/\rm{AU})$. On the left panel, we use an interpolation for the points with $\pi (f - f_n) \Delta T <1$, whereas the right panel, presents the points satisfying $\pi (f - f_n) \Delta T <1$. We adopt $\Delta T = (8, 7, 5, 2, 1, 0.5, 0.5)$ \rm{yrs}. 

\begin{figure*}
\includegraphics[height=160pt,width=500pt,trim = 6mm 1mm 0mm 1mm]{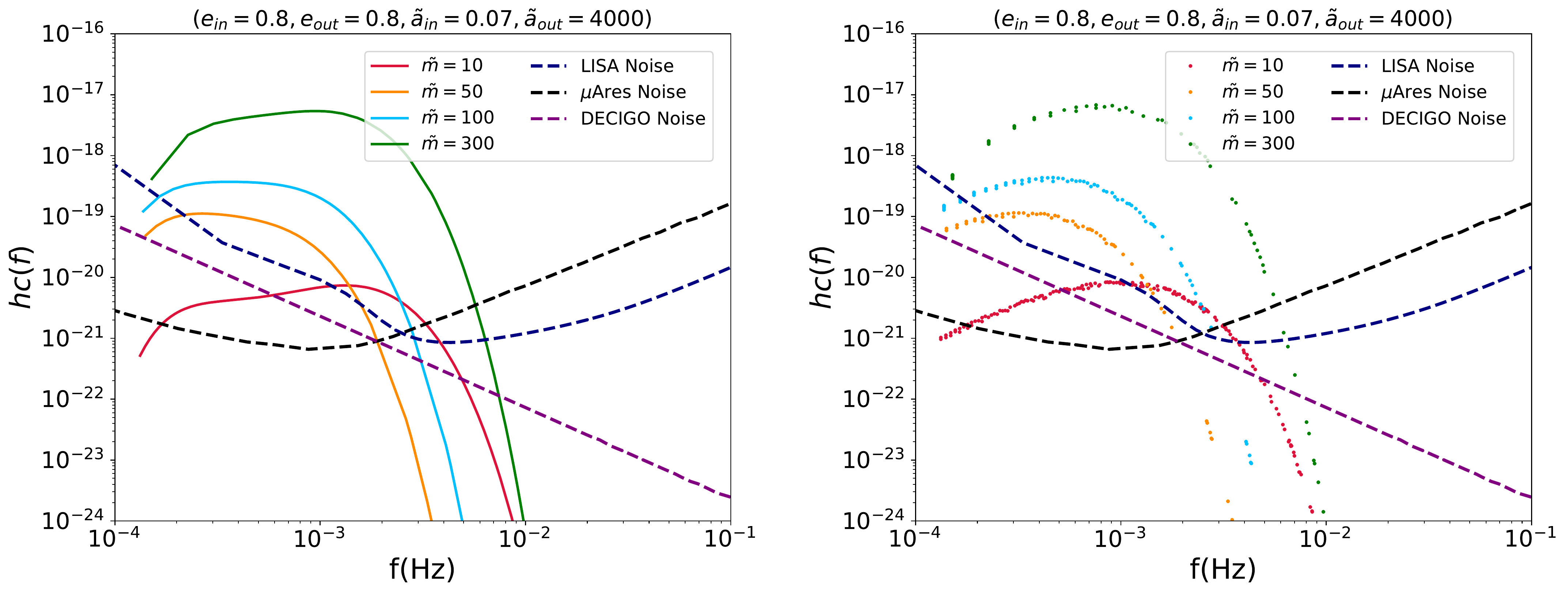}
\caption{GW signal (in color) compared to LISA (dashed blue), $\mu$-Ares (dashed black) and DECIGO (dashed purple) noises for BBHs with individual mass $ \tilde{m} = (10, 50, 100, 300) $. Left: the interpolated GW signal with $\pi (f - f_n) \Delta T <1$. Right: discrete frequencies with the above condition. Signal to Noise ratio for DECIGO are given by \rm{S/N} = (52, 100, 560, 1.44 $\times 10^4$). }
\label{Decigo-com}
\end{figure*}

\begin{figure}
\includegraphics[width=\columnwidth]{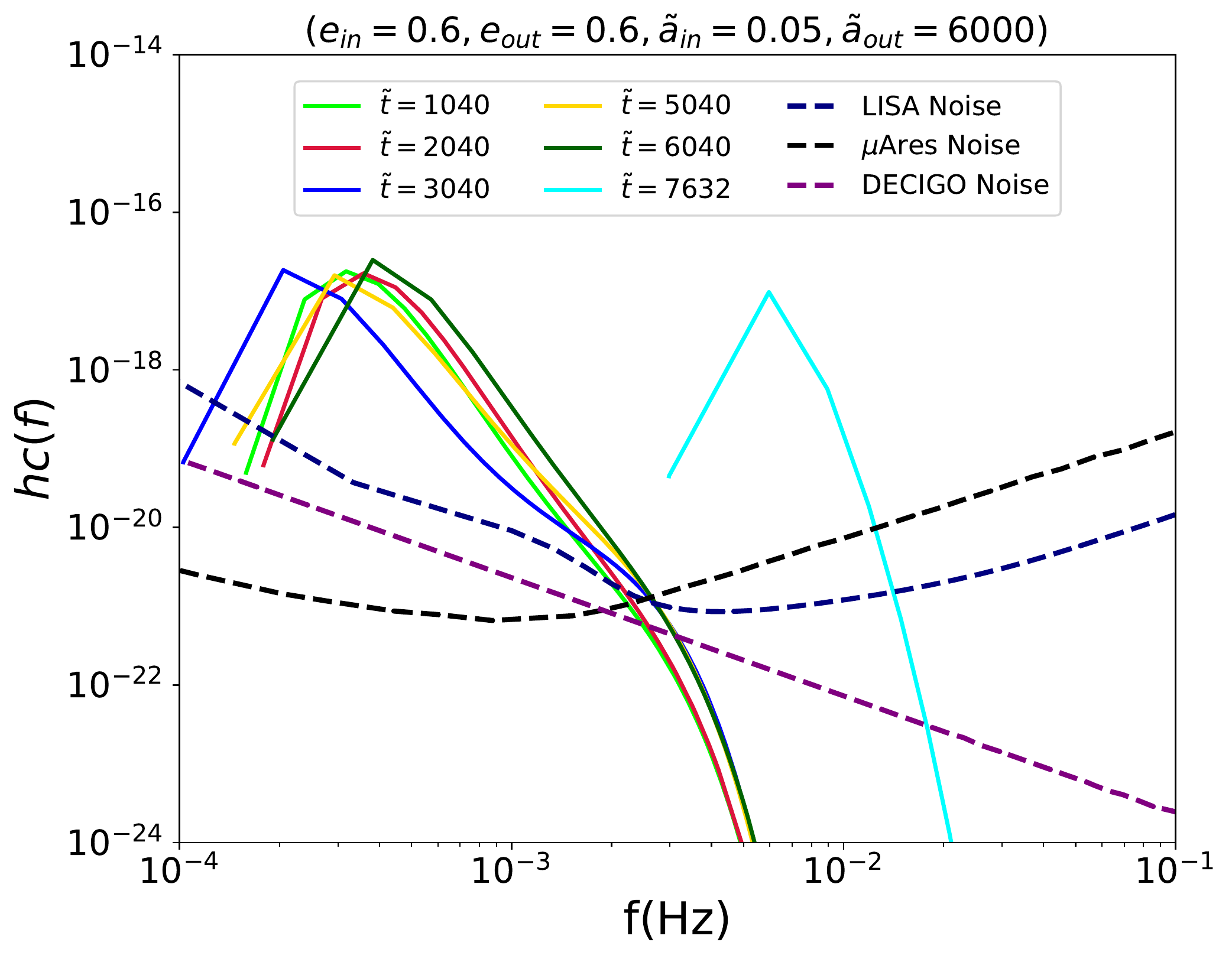}
\caption{The time evolution of the GW signal over a wide range of frequencies. We have taken $\tilde{t} = (1040, 2040, 3040, 5040, 6040, 7632)$, where $\tilde{t} \equiv (t/\rm{yr})$, and $\Delta T = (9, 8, 6, 5, 4, 0.1) \rm{month}$.}
\label{Time-Evolution-LL}
\end{figure}

\subsubsection{Probing GW with the $\mu$-Ares detector}
Since the LISA noise increases at frequencies below \rm{milli-Hz}, it is advantageous to use the newly proposed $\mu$-Ares detectors for detecting GW signals in the frequency range from $\mu$Hz to \rm{milli-H}z. Figure \ref{mass-points-Mu-Ares} presents the characteristic strength of GW at those frequencies. For BH masses with $ \tilde{m} = (10, 50, 100, 300) $. The plot is shown for $\tilde{a}_{in} = 0.1, \tilde{a}_{out} = 4000, e_{in} = 0.8 , e_{out} = 0.7$. Comparing Figure \ref{mass-points-Mu-Ares} with Figure \ref{mass-points}, reveals that the strength of GW for $\tilde{m} = 10$ is enhanced for larger value of inner semi-major axes. This is due to the Kozai-Lidov oscillations which boost the GW signal above the noise toward larger frequencies. Here we consider $\Delta T = (10, 8, 5, 2)$ \rm{yrs}. 

\subsubsection{Probing GW with Decihertz Observatories (DOs)}
Finally, we consider decihertz frequencies, $f \sim (0.01-1)$ \rm{Hz}, which are particularly suitable to IMBHs \citep{2019arXiv190811375A}. There are currently different proposed technologies for probing the GWs within this frequency range. They include DO-optimal, DO-conservative, Advanced Laser Interferometer Antenna (ALIA) and decihertz Interferometer GW observatory (DECIGO) (see, e.g \cite{2019arXiv190811375A} and Refs. therein). In our analysis below, we focus on DECIGO. Figure \ref{Decigo-com} presents the characteristic GW strain against various detectors including the LISA (blue line), $\mu$Ares (black line) and DECIGO (purple line) for $\Delta T = (10, 8, 5, 3)$ \rm{yrs}.

There are clearly overlapping regions in frequency for these three observatories. This is particularly helpful in removing
degeneracies between various parameters in the system. Multi-wavelength spectroscopy of GW can provide novel information about the parameters of the BBHs that are otherwise degenerate, and could be potentially used as a way to discover IMBHs in M87.  

\subsubsection{Time evolution of GW amplitude }
Having presented the frequency evolution of the GW signal for a fixed time, we next study how the signal changes with time. 
In Figure \ref{Time-Evolution-LL}, we draw the evolved GW in a wide range of frequencies. Here we adopt $\tilde{m} = 300$, $e_{in} = 0.6, e_{out} = 0.6, \tilde{a}_{in} = 0.05, \tilde{a}_{out} = 6000$. We consider $\tilde{t} = (1040, 2040, 3040, 5040, 6040, 7632)$ for $\Delta T = (9, 8, 6, 5, 4, 0.1)~\rm{month}$. In this example, increasing $\Delta T$ mostly affects the detectability of the GW signal for the DECIGO observatory.
 
\subsubsection{Impact of orbital parameters in GW amplitude}

In our simulations, we fix some of the orbital parameters such as the arguments of pericenter for the inner and outer binaries (taken to be $0 ^{\circ}$), longitude of ascending node for inner and outer binaries (chosen to be $0 ^{\circ}$) and mutual inclination (taken to be $90 ^{\circ}$). We have also taken the BHs to have zero spins, but allowed rest of the parameters to vary. This includes the inner and outer semi-major axes and eccentricities. We noticed that changing the outer semi-major axis has very minor impact on the results as long as the BBH is far from the tidal disruption distance, $a_{out} \gg a_{in} (M_{SMBH}/M_{BH})^{1/3}$. Similarly, changing $e_{out}$ does not affect the signal significantly. On the other hand, changing $a_{in}$ and especially $e_{in}$ affect the strength of the signal dramatically. Owing to the importance of these parameters, we consider their effect on the GW signal for multiple detectors. 

\begin{figure}
\includegraphics[width=\columnwidth]{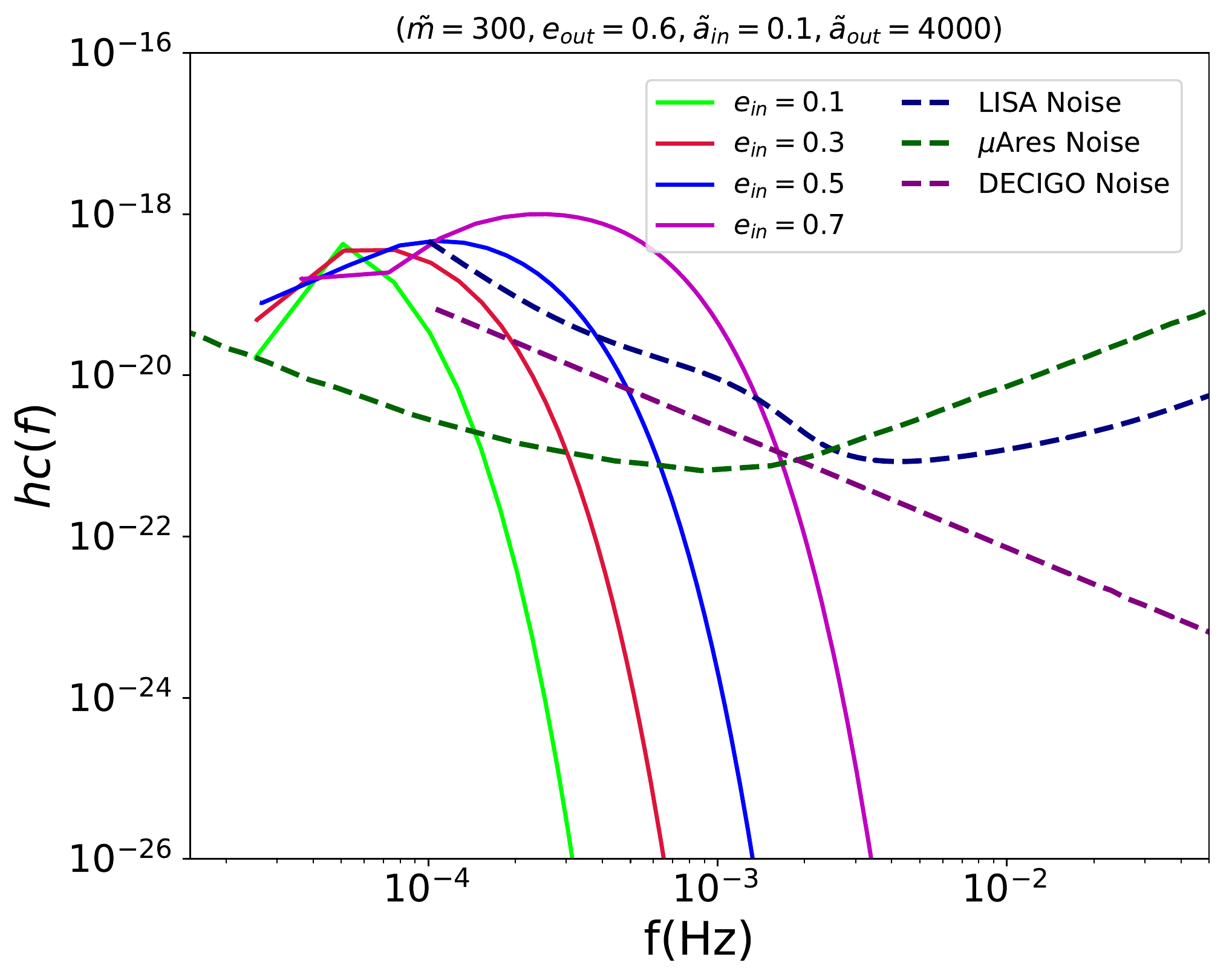}
\caption{The dependence of the GW signal on $e_{in} = (0.1, 0.3, 0.5, 0.7)$, $\tilde{m}= 300, e_{out} = 0.6, \tilde{a}_{in} = 0.1, \tilde{a}_{out} = 4000$, and $\Delta T = 2 \rm{yr}s$.}
\label{Signal-change-inner-e}
\end{figure}

Figure \ref{Signal-change-inner-e} presents the influence of $e_{in}$ on the GW signal, assuming $\tilde{m}= 300, e_{out} = 0.6, \tilde{a}_{in} = 0.1$ and $\tilde{a}_{out} = 4000$. The resulting GW signal could be observed by $\mu$Ares, LISA  or DECIGO detectors. 

\begin{figure}
\includegraphics[width=\columnwidth]{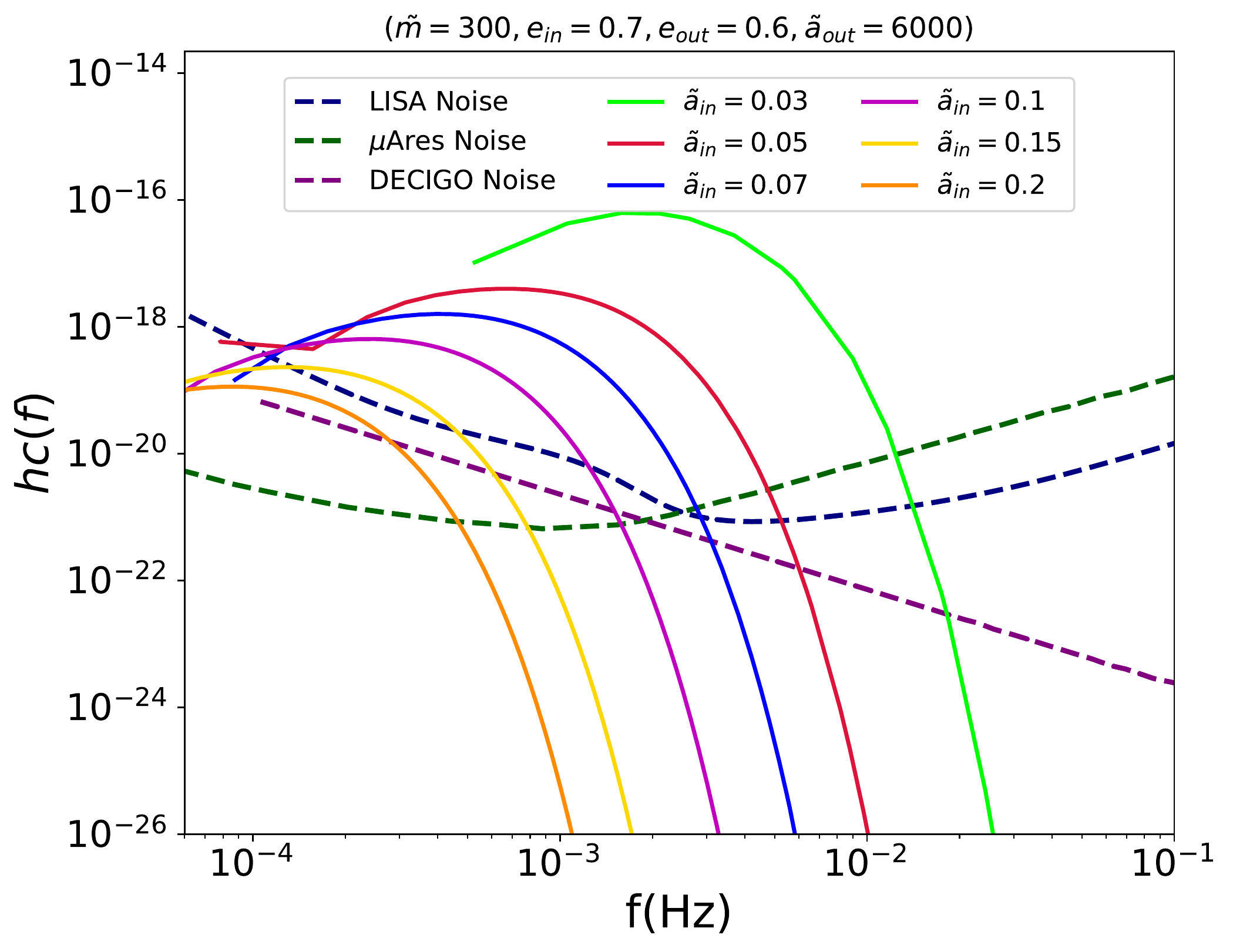}
\caption{The dependence of the GW signal on $\tilde{a}_{in} = (0.03, 0.05, 0.07, 0.1, 0.15, 0.2)$, for $\tilde{m} = 300, e_{in} = 0.7, e_{out} = 0.6, \tilde{a}_{out} = 6000$, and $\Delta T = 2 \rm{yrs}$.}
\label{Signal-change-inner-a}
\end{figure}

Finally, in Figure \ref{Signal-change-inner-a} we examine the impact of changing $\tilde{a}_{in} = (0.03, 0.05, 0.07, 0.1, 0.15, 0.2)$ on the detectability of GW signal. Here we adopt $\tilde{m} = 300, e_{in} = 0.7, e_{out} = 0.6, \tilde{a}_{out} = 6000$. The plot shows that changing $a_{in} $ only affects slightly the GW signal. 
\begin{table}
	\centering
	\caption{Frequency band for different GW observatories.}
	\label{Tab:SN}
	\begin{tabular}{lccr} 
		\hline
		\hline
		 Detector & 
        $f_{min}$(\rm{Hz}) &
        $f_{max}$(\rm{Hz})   \\
		\hline
       $\mu$Ares & 
        $max(10^{-6}, f_m)$
        & $10^{-3}$
        \\
        \hline
       LISA 
       & $f_m$
       &  $10^{-1}$
      \\
       	\hline
       DECIGO 
       &   $10^{-3}$
       & 10
       & \\
       \hline
        \hline
	\end{tabular}
\end{table}
\section{Detectability of GW signal}
\label{Detection}
\begin{table*}
	\centering
	\caption{Signal to Noise (\rm{S/N}) ratio for different BH masses and with $\tilde{a}_{in} = 0.1, \tilde{a}_{out} = 4000, e_{in} = 0.8, e_{out} = 0.7$.}
	\label{SN-example}
	\begin{tabular}{lccccr} 
		\hline
		 BH mass $(M_{\odot})$ & 
        $\mu$Ares &
        LISA &
        DECIGO &
           \\
		\hline
       ~~~~~10 & 
        3.6 &
        9.4 &
        38.5 &
        \\
        \hline
        ~~~~~50 & 
        41.1 &
        1.2 &
        0.3 &
        \\
        \hline
        ~~~~~100 & 
        148 &
        4.8 &
        2.6 &
        \\
        \hline
        ~~~~~300 & 
        $1.7 \times 10^{3}$ &
        86 &
        163 &
        \\
        \hline
         ~~~~~500 & 
        $ 4.3 \times 10^{3}$ &
        $ 3.2 \times 10^{2}$ &
        793 &
        \\
        \hline
        ~~~~~700 & 
        $ 5.5 \times 10^{3}$ &
        $ 5.45 \times 10^{2}$ &
       $ 1.4 \times 10^{3}$ &
        \\
        \hline
         ~~~~~1000 & 
        $ 1.6 \times 10^{4}$ &
        $ 2.5 \times 10^{3}$ &
       $ 6.7 \times 10^{3}$ &
        \\
        \hline
	\end{tabular}
\end{table*}

Next, we consider the detectability of GW signals for some of the examples above. We compute the signal to noise ratio (\rm{S/N}) using Eq. (\ref{signal-noise}), and label a GW signal as detectable if $S/N \geq 10$. The exact lower limit in $S/N$ is not fixed. Different values are used in the literature. For example \cite{2019ApJ...875L..31H} used $S/N \geq 5$ as the criteria of detection. Here we use a more conservative criteria of 10 as our detection limit. 
Since different GW detectors are focused on different frequency bands, we need to define the frequency bands for different detectors using Eq. (\ref{signal-noise}). Table \ref{Tab:SN} presents the frequency ranges for different GW detectors. Although for most cases we take the universal upper and lower limits in the integrals, the lower limit for $\mu$Ares depends on the minimum value of the frequency, which could be slightly larger than $10^{-6} \rm{Hz}$. Owing to this, we take the maximum value between $10^{-6} \rm{Hz}$ and $f_m$, defined to be the minimum value of the frequency of GW signal. Likewise, for the LISA experiment, we take the minimum value of the frequency of signal as the lower limit of the integral.

In addition, since the GW frequency is chirped towards the merger, the computation of the \rm{SN} also depends on the approximate point in the evolution of the system. In other words, the time evolution of the system affects the GW signal and so the \rm{S/N}. Therefore, lower signal to noise may evolve with time and get enhanced. With this in our mind, in the following we present \rm{S/N} ratio for one set of the examples. Table \ref{SN-example} presents the signal to noise ratio for the case with $\tilde{m} \equiv m/M_{\odot} = (10, 50, 100, 300, 500, 700, 1000)$ and with $\tilde{a}_{in} = 0.1, \tilde{a}_{out} = 4000, e_{in} = 0.8, e_{out} = 0.7$, and $\Delta T = (9, 8, 5, 3, 2, 1, 1, 1) \rm{yr}$. 
The GW signal is evaluated at the same time for all cases. Since the dynamical evolution of different BHs differ depending on their masses, the orbital and peak freak frequency, 
\ba 
\label{fpeak}
f_{p} = 2 f_{orb} \frac{\left(1 + e_{in} \right)^{1.1954}}{\left( 1 - e_{in}^2 \right)^{3/2}},
\ea
is different in these examples. Therefore over the time, the location of signal  changes for different BH masses and we may put them in the same location if we look at individual ones at different times.

\begin{figure}
\includegraphics[width=\columnwidth]{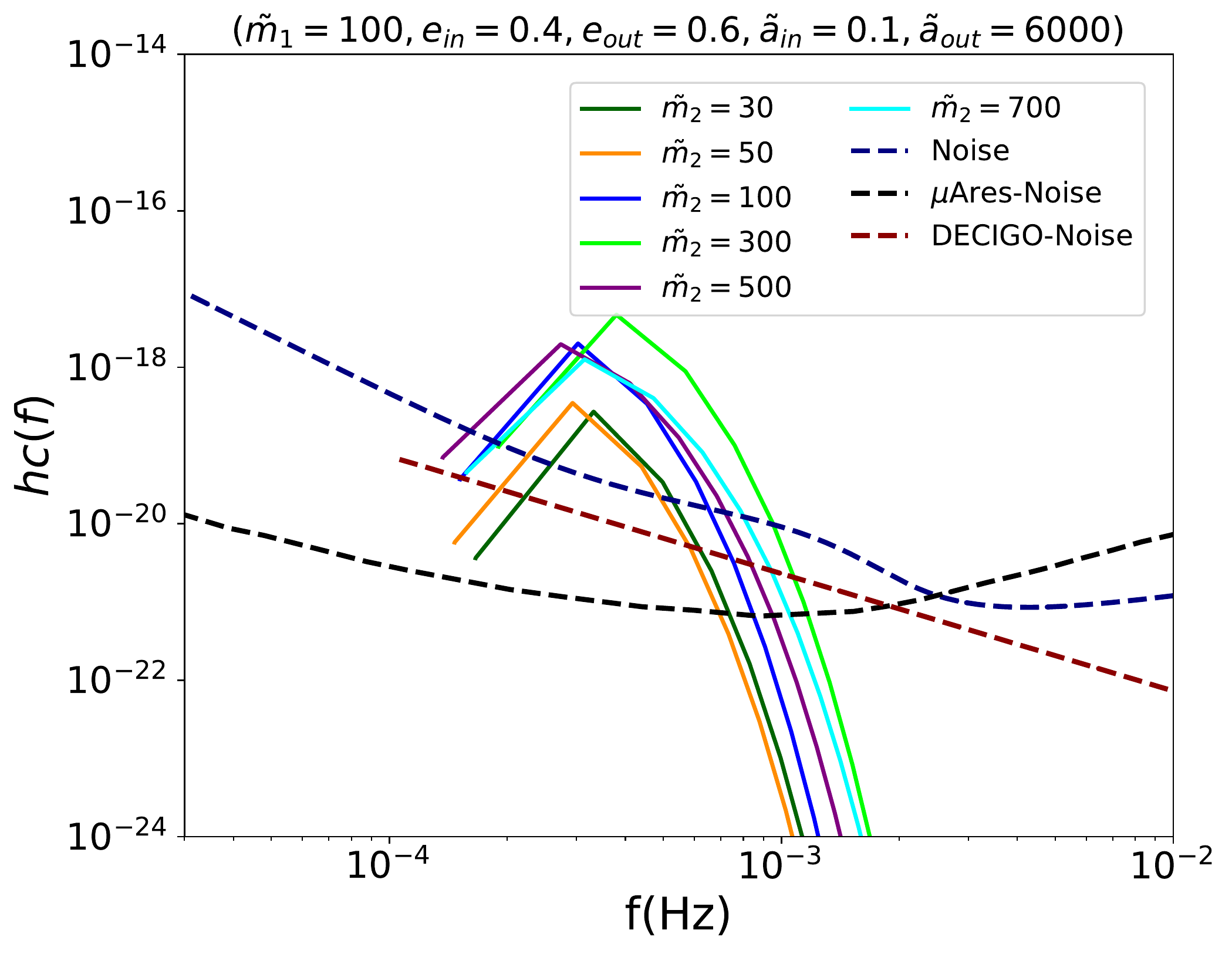}
\caption{GW signal for non-equal BH masses. Here we adopt $m_{1} = 100 M_{\odot}$ and consider different values for its companion mass in the range $m_{2} = (30, 50, 100, 300, 500, 700) M_{\odot}$. The remaining orbital parameters are taken to be $a_{in} = 0.1 \rm{AU}, e_{in} = 0.4, e_{out} = 0.6, a_{out} = 6000 \rm{AU}$.}
\label{change-mass}
\end{figure}

\section{Non-Equal mass BHs}
\label{non-equal}
So far, we only focused on the BBHs with equal masses. Here we briefly consider the case with non-equal BH masses. As a test example, we fix the mass of one of BHs to be $m_{1} = 100 M_{\odot}$ and change the mass of its companion in the range $m_{2} = (30, 50, 100, 300, 500, 700) M_{\odot}$. The rest of the orbital parameters are taken to be $a_{in} = 0.1 \rm{AU}, e_{in} = 0.4, e_{out} = 0.6, a_{out} = 6000 \rm{AU}$. Figure \ref{change-mass} presents the GW signal for this regime. 

Unlike the examples in Sec. \ref{Detection}, we 
evaluate the GW signal at the time for which the peak frequency (defined in Eq. \ref{fpeak}) to be around $ f_{p} = 3 \times 10^{-4} \rm{Hz}$. 
This makes the GW signal behave very similarly. Therefore the peak frequency is a key parameter in our system and leads to an almost universal behavior at different BH masses. 
In closing, we note that each IMBH may carry a cluster of stellar-mass BHs around it, enhancing the rate of detectable GW signals from its vicinity. 

\begin{figure*}
\includegraphics[height=200pt,width=510pt,trim = 6mm 1mm 0mm 1mm]{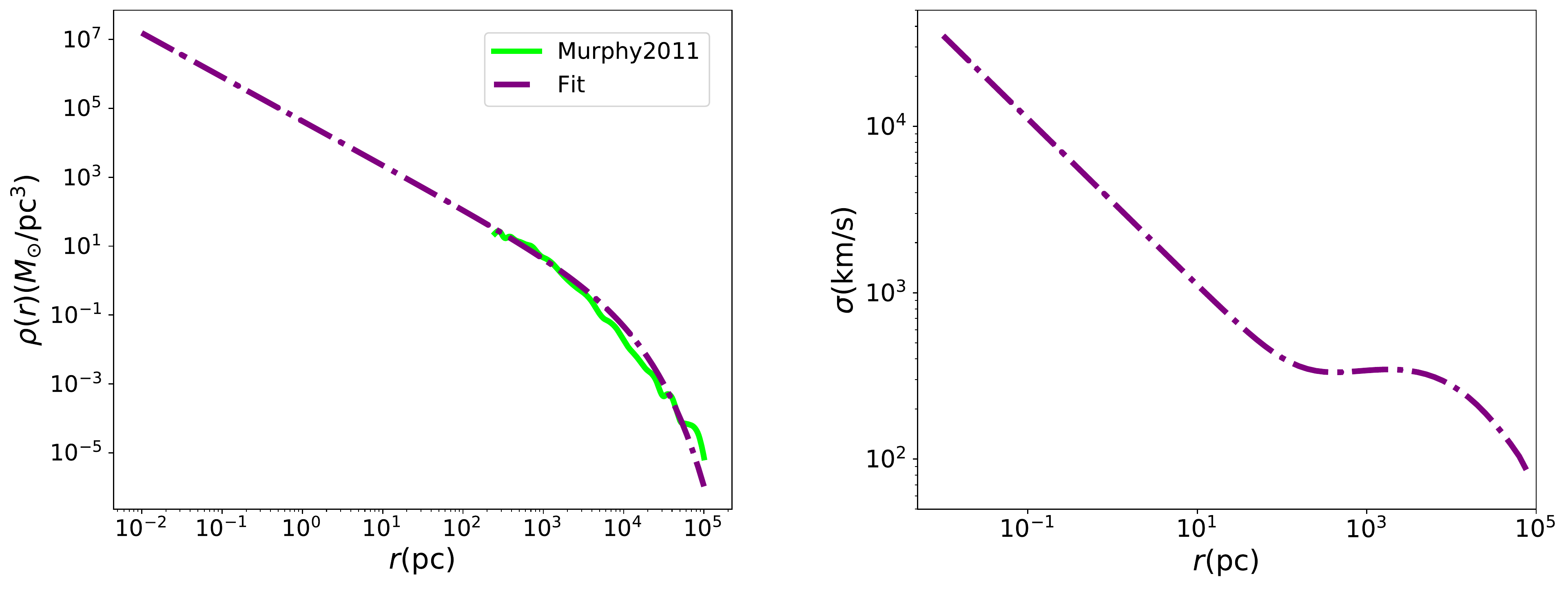}
\caption{\textbf{Density profile (left) and velocity dispersion (right) of M87. }}
\label{density87}
\end{figure*}

\section{Interaction between BBH and flybys}
\label{Collision-Rate}
So far we considered the impact of the SMBH on the evolution of BBHs at M87 and its imprint on gravitational waves. We ignored the impact of any flybys in the evolution of system. Since M87 is surrounded by a stellar cluster, there might be some flyby stars that reach BBHs. It is therefore intriguing to have an order of magnitude estimation on the rate of collisions leaving its comprehensive study for a future paper. 

The rate of different collisions and dynamical interactions in galactic nuclei and globular clusters have been considered in \cite{2016MNRAS.463.1605L,2014MNRAS.444...29L,2018MNRAS.474.5672L}. 
Following \cite{2018MNRAS.474.5672L}, we compute the rate of interaction between an inner BBH and a cloud of stars belonging to the stellar cluster around M87. We ignore the impact of the stellar disk and focus on the dispersion supported spherical stellar component of M87. 
The rate of collisions between the inner BBH and stars is given by \cite{2018MNRAS.474.5672L}, 

\ba 
\label{collision-rate}
\Gamma \simeq \rho(r) \left( \frac{R^2}{M} \right) \left( 1 + \left( \frac{v_e}{\sigma(r)} \right)^2 \right) \sigma(r),
\ea
where $\rho(r)$ and $\sigma(r)$ describe to the mass density of dispersion supported stellar component and velocity dispersion next to SMBH with $r$ referring to the distance from the central SMBH. Here, $v_e = \left( \frac{2G M} {R} \right)^{1/2}$ describes the escape velocity from the BBH. Finally, in the above formula we take $M$ and $R$ to be the mass of BBH as well as the semi-major of orbit, respectively. 

For the density profile, $\rho(r)$, we use Figure 9 of \cite{2011ApJ...729..129M} to infer the total enclosed mass of stars at every radius $r$ from M87. Density profile can then be computed from the derivative of the enclosed mass. We fit a non-linear function to the density profile and extend it to smaller radii above the data resolution. The left panel in Figure \ref{density87} presents the density profile of M87 for \cite{2011ApJ...729..129M} and our non-linear fit.

Using the above extended mass/density profiles, we next compute the velocity dispersion as,

\ba 
\label{Velocity}
\sigma^2(r)= \left( \frac{G}{\rho(r)} \right) \int_{r}^{\infty} \left( \frac{\rho(r') }{r'^2} \right) \bigg{(}M_{SMBH} + M_{\star}(r') \bigg{)} dr',
\ea
The right panel in Figure \ref{density87} displays the radial profile of the velocity dispersion. 

Plugging the above results back into Eq. (\ref{collision-rate}), we find that the collision rate of stars to BBH, $\Gamma$, depends on the BBH mass and semi-major as well as the distance from the center. To estimate $\Gamma$, we fix r = 6000 \rm{AU} = 0.03 \rm{pc} $\simeq$  46.6 $R_g$ and compute the collision rate as a function of the BBH mass and inner semi-major axes.

Figure \ref{collision-rate} displays the behavior of $\log_{10}{(\Gamma/\rm{Gyr}^{-1})}$ as a function of the logarithm of BBH mass, $\log_{10}{(M/M_{\odot})}$ and logarithm of semi-major axes $\log_{10}{(R/\rm{AU})}$. As expected, the rate enhances for larger/smaller BBH semi-major/mass, respectively. 


\begin{figure}
\includegraphics[height=190pt,width=250pt,trim = 6mm 1mm 0mm 1mm]{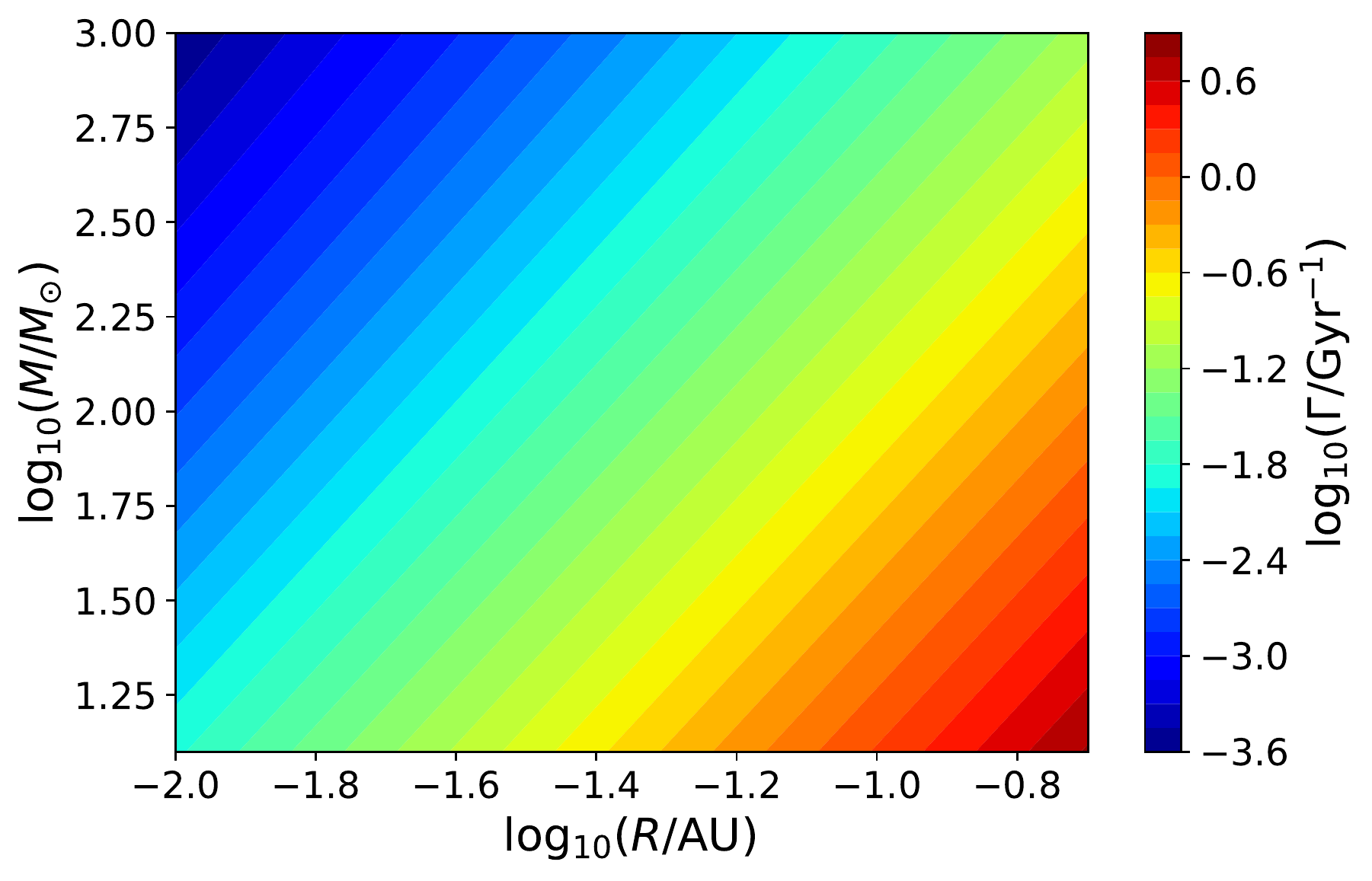}
\caption{Collision rate of inner BBH and the flyby stars. $M$ and $R$ refer to the mass and semi-major axes of BBH, respectively.}
\label{collision-rate}
\end{figure}

\begin{figure*}
\includegraphics[height=160pt,width=500pt,trim = 6mm 1mm 0mm 1mm]{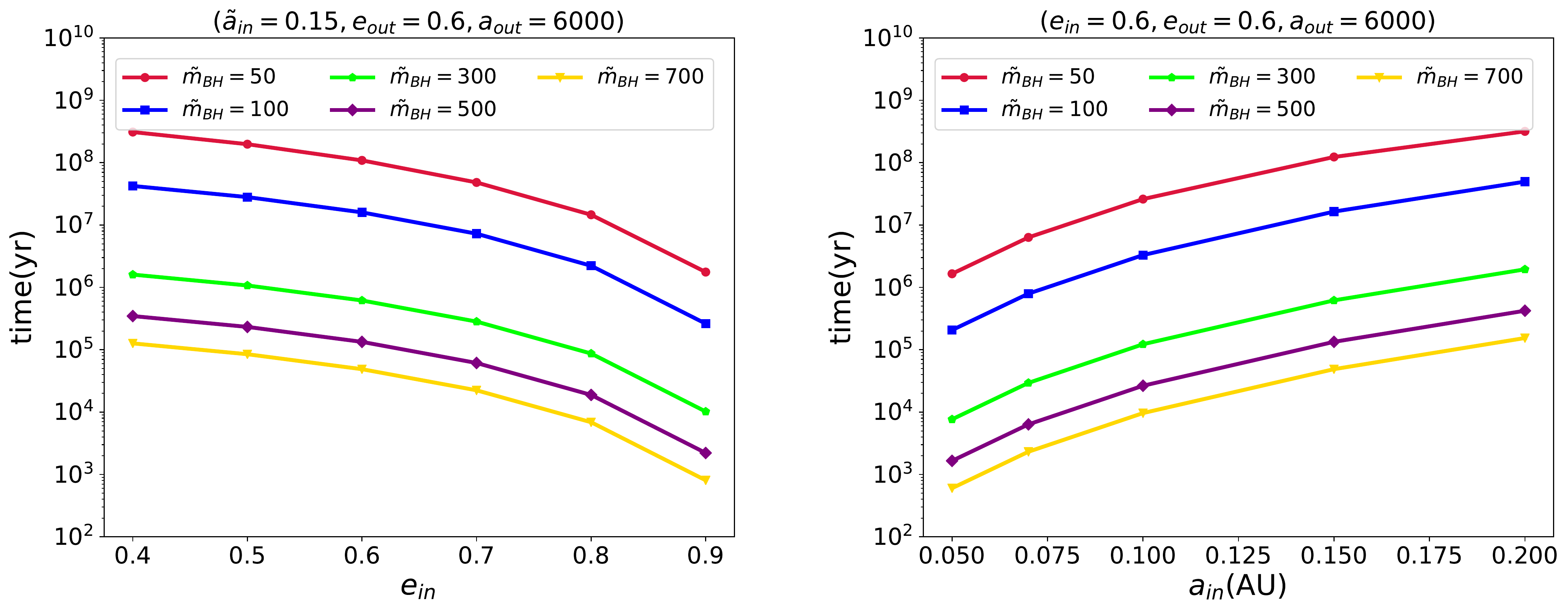}
\caption{The lifetime of the BHs in the inner binary as a function of $e_{in}$ (left) and $a_{in}$ (right) and for different values of the BH masses. }
\label{LifeTime}
\end{figure*}

\section{Lifetime of ORBITS}
\label{Lifetime}
Finally, we consider the lifetime of BBHs orbits under consideration. By the lifetime we mean the time that it takes for an inner binary with given initial parameters to merge. The above timescale is not meant to represent the detectability window. In \cite{2020JCAP...02..021E} we computed the rate for the inspiral of stellar mass BHs around the SMBH. Nonetheless the above timescale provides a sense about the lifetime of a given orbit under consideration.
Binaries with very short lifetime do not survive. Their replenishment requires some secondary process after the merger such as gas supply \citep{2019ApJ...878...85S}. On the other hand, binaries with long lifetime have a better chance of being seen in the real observations provided that they are not evaporated or tidally disrupted by the close flybys. Yet, since we are mostly interested in IMBHs with relatively large masses, the chance for a BBH to get destroyed by some dynamical interactions is not expected to be high. 
A comprehensive analysis of the rate of these binaries including the above effects goes beyond the scope of this paper. 

Figure \ref{LifeTime} presents the lifetime of BBHs for different values of their masses and as a function of $e_{in}$ and $a_{in}$. For simplicity, we only focus on equal mass BHs.
From the plot it is clear that the lifetime of the BBHs is a strong function of the orbital parameters as well as the BH masses. As expected, the lifetime increases by decreasing the values of $e_{in}$ and increasing the value of $a_{in}$.

\section{Conclusions}
\label{Conc}
Multi-Wavelength GW detectors can monitor the spectrum of GW signals
from the center of M87 over a wide range of frequencies and orbital parameters. GW spectroscopy enables one to reproduce the shape of the GW signal and get novel information about the physical process behind such signals. 
Focusing on triple systems in M87 made of an inner binary BHs with different masses, from stellar to IMBHs, we presented a consistent method for detecting GW signals by integrating over the observation time within the lifetime of different GW detectors. We demonstrated that the frequency peaks from various GW sources can be used to entangle the signal from closer in sources with continuum frequency bands. 
The parameter space where different GW detectors may overlap in their frequency of GWs, could be used as a novel way to break the degeneracy between different orbital parameters. 

We estimated the rate of collision between the BBHs in M87 and the flyby stars belong to the stellar cluster at the central region of M87. Though the collision rate depends on the mass and the semi-major axes of inner binary, our results imply that the rate is smaller than 10 $\rm{Gyr}^{-1}$ for the wide range of parameters considered here.

A more detailed numerical simulation over a wider time range including all of the possible encounters between different BHs is left for a future investigation.

\section*{Acknowledgements}
We thank Bao-Minh Hoang, Bence Kocsis and Smadar Naoz for the thoughtful comments on the manuscript. We are thankful to Fabio Antonini, Evgeni Grishin and Scott A. Hughes for the interesting discussion. We are very grateful to an anonymous referee for their  constructive comments that improved the quality of this paper.
R.E. acknowledges the support by the Institute for Theory and Computation at the Center for Astrophysics.
This work was also supported in part by the Black Hole Initiative at Harvard University which is funded by grants from the Templeton and Moore foundations. We thank the supercomputer facility at Harvard where most of the simulation work was done.


\bibliographystyle{mnras}
\interlinepenalty=10000
\bibliography{Main}


\bsp	
\label{lastpage}
\end{document}